# Structural Contribution to the Ferroelectric Fatigue in Lead Zirconate Titanate (PZT) Ceramics

*M. Hinterstein[1,2], J. Rouquette[3,*], J. Haines[3], Ph. Papet[3], J. Glaum[1], M. Knapp[2], J. Eckert[4], M. Hoffman[1]*

[1] *School of Materials Science and Engineering, University of New South Wales, NSW, Australia.*

[2] *Institut für Angewandte Materialien – Energiespeichersysteme, Karlsruher Institut für Technologie, 76344 Eggenstein-Leopoldshafen, Germany.*

[3] *Institut Charles Gerhardt UMR CNRS 5253 Equipe $C_2M$, Université Montpellier II, Place Eugène Bataillon, cc1504, 34095 Montpellier cedex 5, France.*

[4] *Institut für Werkstoffwissenschaft, Technische Universität Dresden, 01062 Dresden, Germany.*

AUTHOR EMAIL ADDRESS: Jerome.Rouquette@univ-montp2.fr

**Many ferroelectric devices are based on doped lead zirconate titanate (PZT) ceramics with compositions near the morphotropic phase boundary (MPB), at which the relevant material's properties approach their maximum. Based on a synchrotron X-ray diffraction study of MPB PZT, bulk fatigue is unambiguosly found to arise from a less effective field induced tetragonal-to-monoclinic transformation, at which the degradation of the polarisation flipping is detected by a**

**less intense and more diffuse anomaly in the atomic displacement parameter of lead. The time dependance of the ferroelectric response on a structural level down to 250 µs confirms this interpretation in the time scale of the piezoelectric strain response.**

While the puzzle of thin film fatigue has been solved to a large degree in the last two decades [1-45], the mechanisms for bulk actuator fatigue remains still unsolved and particularly the structural contribution to the ferroelectric fatigue in bulk materials is not well understood in spite of intensive study[6]. Ferroelectric fatigue is either proposed to be caused by different mechanisms on the macroscopic or mesoscopic length scales, such as the formation of interfacial layers between the metal electrodes and the ferroelectric material[7, 8], 90-degree domains[5] and resultant strains, electromigration of oxygen vacancies to form extended defects capable of pinning domains[9, 10], interfacial nucleation inhibition mechanisms induced by charge injection[11, 12], macro- and/or micro-cracking[13], etc., but to the best of our knowledge, description for fatigue on the crystal structure level has never been proposed for bulk samples. In the literature, X-ray diffraction studies of fatigue as a function of the electric field has focused on microstructural effects, which are generally probed at $\psi$=0° between the incident X-ray beam and the electric field vector, as the large piezoelectric strain does not permit structure refinement to be performed. PZT ceramics with compositions at the morphotropic phase boundary (MPB)[14] between the tetragonal (*P*4*mm*) and rhombohedral (*R*3*m*) phase fields exhibit optimal ferroelectric properties for a large number of technological applications (sensors and actuators, micromechanical systems, and high frequency devices). We recently reported an *in situ* description of the origin of the ferroelectric properties as a function of the applied electric field *E* based on a synchrotron X-ray diffraction study[15] at $\psi$=45° between the incident X-ray beam and the electric field vector to minimize microstructural effects. The monoclinic (pseudo-rhombohedral)/ tetragonal phase ratio was found to increase with electric field, which strongly supported the hypothesis of Noheda et al., who explained the strong piezoelectric properties of PZT by the presence of a monoclinic phase[16] giving rise to a polarization rotation mechanism[17]. Note that in this study as the rhombohedral model could not correctly reproduce the low-field data, this structural model was abandoned in order to describe the electric field dependence of the structural data using the same model and the controversies between the martensitic theory and the polarization rotation models are clearly beyond the scope of the present study. Additionally, polarization flipping of polar lead atoms could unambiguously be characterized by a maximum in the disorder of lead, i.e. $B_{iso}$(Pb)$_m$, which is obtained from its isotropic atomic displacement parameter linked to the maximum of the entropy for the positive-negative value of the coercive field ($E_c$) in the $P_S$-$E$ hysteresis cycle. Based on this successfull microscopic description of the macroscopic ferro- and piezolectric properties of PZT, we decided to investigate the ferroelectric fatigue in this important technological material by synchrotron X-ray diffraction. In addition, while fatigue is found to be mainly an interfacial effect in thin films, all our experiments are performed on bulk ceramics as are used in actuators, sensors and ultrasonic motors applications. Here, we unambiguosly show the structural difference between a non-fatigued and a fatigued sample which arises from i) a less effective field induced tetragonal-to-monoclinic transformation and ii) a significantly less intense and more diffuse anomaly of $B_{iso}$(Pb)$_m$, which accounts for polarisation flipping. Finally, stroboscopic pump-probe experiments are performed in order to investigate the structuralcontribution to the time dependance of the ferroelectric response down to 250 µs which clearly support and spectacularly confirm our interpretation on the structural description of the ferroelectric fatigue in the time scale of the piezolectric strain response.

X-ray measurements were performed at the high resolution powder diffraction station of the materials science beam line at the Swiss Light Source (SLS, Villigen, Switzerland)[18] using high-energy photons $\lambda$ = 0.443187 Å to study thin silver metallized PZT pellets[19] (thickness ≤ 110 µm) in transmission geometry. Commercially available PIC 151 (PI Ceramics, Lederhose, Germany), which we already used[15] as it presented optimal piezo- and ferroelectric properties properties ($d_{33}$ = 500 pC/N), long operation periods and offers a high degree of reproducibility, was used to compare unfatigued and fatigued samples. The experimental set up can be found elsewhere[15]. The time to measure a complete ferroelectric hysteresis cycle was about 20 min, resulting in a cycling frequency of 0.8 mHz. The kinetics of the field induced response were investigated with a stroboscopic pump-probe setup. In this technique, the reaction is first triggered with an electric field (pump) and then a diffraction pattern is collected at a specific time delay (probe).

During a stroboscopic experiment the sample has to be cycled many times. Special care has to be taken to remain in a cycling range where no fatigue occurs. A normal measurement needs 10 s of exposure time. For bipolar cycling with 50 Hz and a time resolution of 1 ms, collection of the whole cycle yields 20 diffraction patterns. Since the setup at the MS beamline allows the measurement of only one time channel at a time, just 1/20 of the overall X-ray intensity is collected during every measurement. This results in an exposure time of 200 s per pattern and 4000 s for a whole cycle. This means that during the complete measurement, the sample runs through $4000s \cdot 50Hz = 2 \cdot 10^5$ cycles. Based on the to the literature, fatigue can be detected above $10^3$ cycles[20]. However, we took special care to prevent interface effects by using high quality electrodes and we clearly checked that nearly no fatigue could be observed in our material in the range of $10^5$ cycles[19]. In order to assure stable conditions during the stroboscopic measurements, the samples were pre-cycled $10^5$ times with a bipolar electric field of 2.0 kV/mm and a frequency of 50 Hz. In order to investigate the fatigue mechanisms, the samples were bipolar fatigued with $10^7$ cycles.

Fig. 1 shows the difference in the physical (Fig. 1a and 1b) and structural behavior (Fig. 1c and 1d) for an unfatigued and a fatigued sample. As already explained in the introduction, the ferroelectric fatigue is characterized by the loss of the switchable remanent polarization and strain as a function of the number of switching cycles. Figs. 1c and 1d show the electric field dependence using a cycling frequency of 0.8 mHz of the pseudocubic $200_c$ reflection, which was found to be quantitatively representative of the relative abundance of the monoclinic (pseudo-rhombohedral) and tetragonal phases[15]. It is consistent with a decrease of the tetragonal intensity, while the monoclinic intensity in between the tetragonal reflections increases with electric field. However, one can clearly observe that the field induced tetragonal-to-monoclinic transformation is not as effective for the fatigued sample. This is supported by the information obtained by Rietveld refinements of the synchrotron X-ray diffraction patterns (Table 1), Fig. 2. Fig. 2a and 2b shows typical high quality refinements obtained for unfatigued and fatigued PZT respectively. All the data, i.e. 24 hysteresis cycles or 2880 data sets for the unfatigued sample and 10 hysteresis cyles or 1200 data sets for the fatigued sample respectively, were refined with the same monoclinic-tetragonal two-phase model described elsewhere[15]. One can first note that the monoclinic-tetragonal phase fractions obtained on Fig. 3a and 3b for the unfatigued and fatigued samples respectively are consistent with a butterfly shape, which is the characteristic macroscopic signature of these materials. Additionally, the information obtained from these measurements does not exhibit dispersion and is perfectly reproducible. The edges of the butterfly shape, i.e. about 1 kV/mm, which corresponds to the increase in the monoclinic phase fraction clearly defines the coercive field as already reported[15] and this value does not change under cycling either for the unfatigued or for the fatigued samples in agreement with macroscopic measurements obtained in Fig. 1a and 1b. The only large difference between the unfatigued and the fatigued samples is therefore linked to the tetragonal-to-monoclinic transformation, which is roughly shifted by 10% towards the tetragonal phase for the fatigued sample. In agreement with Noheda et al.[16], we attributed the strong piezoelectric response of PZT to the field-induced tetragonal-to-monoclinic transition. Therefore the 10%-phase ratio shift towards the tetragonal phase for the fatigued sample is consistent with the degradation of the macroscopic properties observed on Fig. 1a and 1b.

Furthermore, Fig. 3c and 3d show the electric field dependence of the mean square displacement of the lead atom [$B_{iso}(Pb)_{\mathrm{m}}$] in the monoclinic phase for the unfatigued and the fatigued sample respectively; the monoclinic phase was chosen as it is the most representative, particularly in the high electric field region. $B_{iso}(Pb)_{\mathrm{m}}$ is associated with the disorder of the considered atom[21,15] and is therefore very sensitive to any structural change. $B_{iso}(Pb)_{\mathrm{m}}$ exhibits two anomalies at about 1kV/mm and -1kV/mm, which are consistent with the value of the coercive field. Additionally, these anomalies are always linked to a change in sign of the spontaneous polarization [(1) and (3) in Fig. 1a] and were therefore associated to the polarization flipping of the lead atoms[15]. This means that the polarization flipping can unambiguously be detected by our structural study. One has to keep in mind that lead atoms are located in non centrosymmetric positions, i.e. they have polar displacements in the PZT structure. It is, therefore, reasonable to associate the polarization flipping, carried out by these polar atoms with a maximum in the disorder parameter of lead, i.e. *Biso* (Pb) that characterizes the maximum of the entropy.

Finally in order to investigate the kinetics of the field induced phase transition, we performed stroboscopic pump-probe experiments with a frequency of 50 Hz. Fig. 4 shows the characteristic $200_c$ reflection for the unfatigued and fatigued sample respectively as a function of the applied electric field for rectangular on-off stroboscopic pump-probe measurements. It clearly demonstrates that the field induced structural response has a time scale in the range of hundreds of microseconds as observed based on macroscopic strain measurements[22-24]. After 1.5 ms, no further structural changes were observed. More importantly, it provides clear evidence for the structural contribution to the ferroelectric fatigue over a shorter timescale. The unfatigued sample exhibits a perfect electric field induced tetragonal-to-monoclinic phase transition, while the same feature is clearly incomplete for the fatigued sample.

In conclusion, in this study the ferroelectric fatigue of commercial PZT ceramics was investigated by *in situ* X-ray synchrotron diffraction experiments as a function of the applied electric field. Using the $\psi$=45° scattering geometry between the incident X-ray beam and the electric field vector previously described to minimize microstructural effects, we unambiguously characterized contribution of the crystal structure to the fatigue based on a reduced degree of tetragonal-to-monoclinic transformation, which accounts for the reduction of the piezoelectric/ferroelectric efficiency. Loss of the switchable remanent polarization could also be evidenced by a significantly less intense and more diffuse anomaly of $B_{iso}$(Pb)$_m$ for the fatigued sample, which accounts for polarisation flipping. We were also able to characterize structurally the field induced response on hundred-microseconds time scale as observed based on macroscopic strain measurements[22, 23], which confirms without any doubt the structural description of the ferroelectric fatigue on a shorter timescale. These results will be useful to design and optimize high performance, fatigue free materials.

This work was performed at the Swiss Light Source, Paul Scherrer Institute, Villigen, Switzerland. The authors would like to thank F. Gozzo and A. Cervellino for the beamline support and A. Bergamaschi for the competent help with the MYTHEN detector system. The research leading to these results has received funding from the BMBF (*Bundesministerium für Bildung und Forschung*) project grant no. 05K13PSA and from the *Alexander von Humboldt Foundation* project grant no. 3.1 - DEU/1150785. The authors appreciate the financial support of the German Research Foundation (DFG) through the *Sonderforschungsbereich* 595 'Electric fatigue in functional materials' and the German Crystallographic Society (DGK).

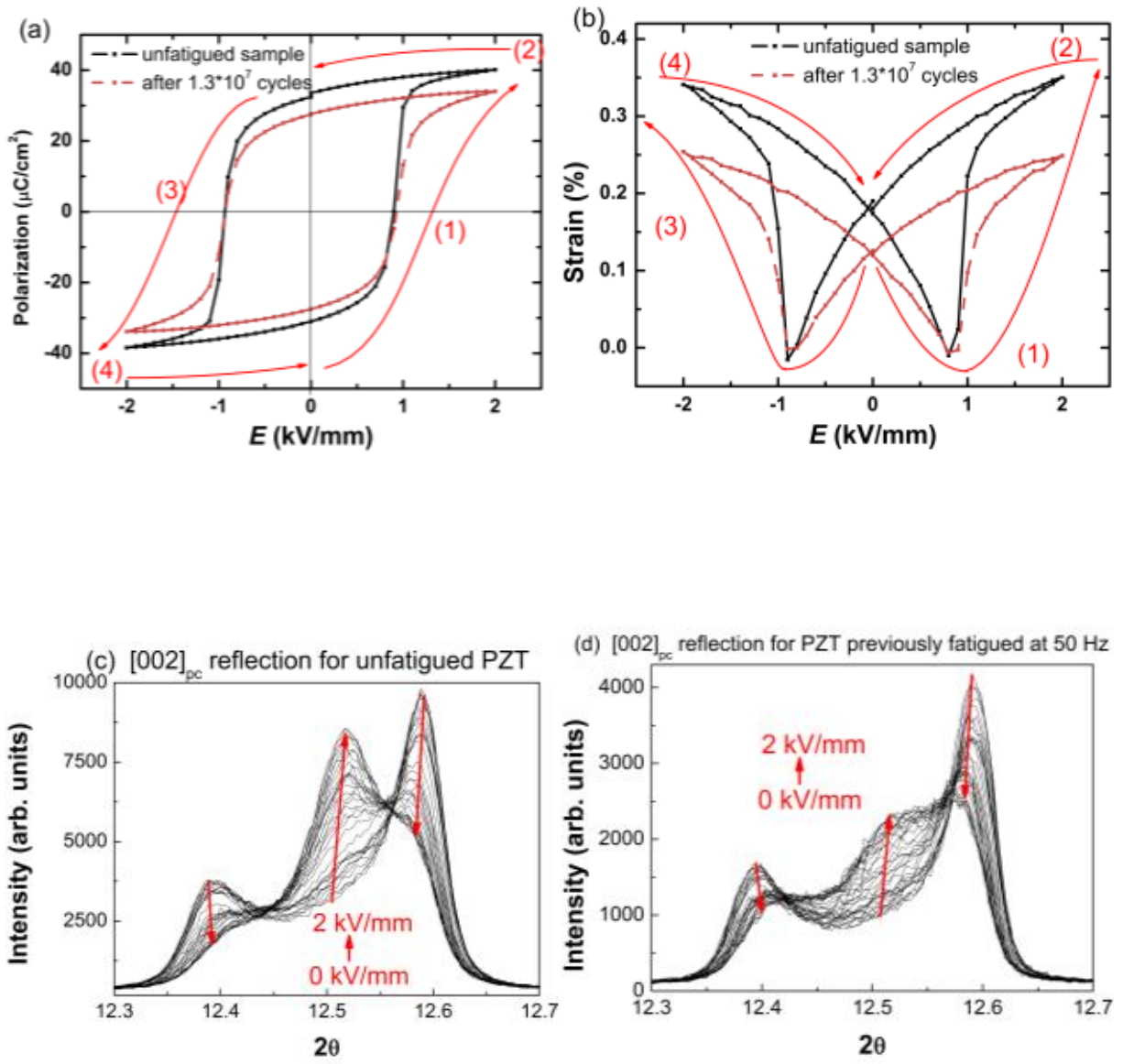

FIG. 1. Difference in the physical and structural behavior for an unfatigued and a fatigued PZT. (a) Cycling dependence of the spontaneous polarization and of (b) field induced-strain; hysteresis cycle (HC) electric field values range (1-4). Influence of the applied electric field on the $002_c$ reflection for (c) unfatigued and (d) a fatigued sample.

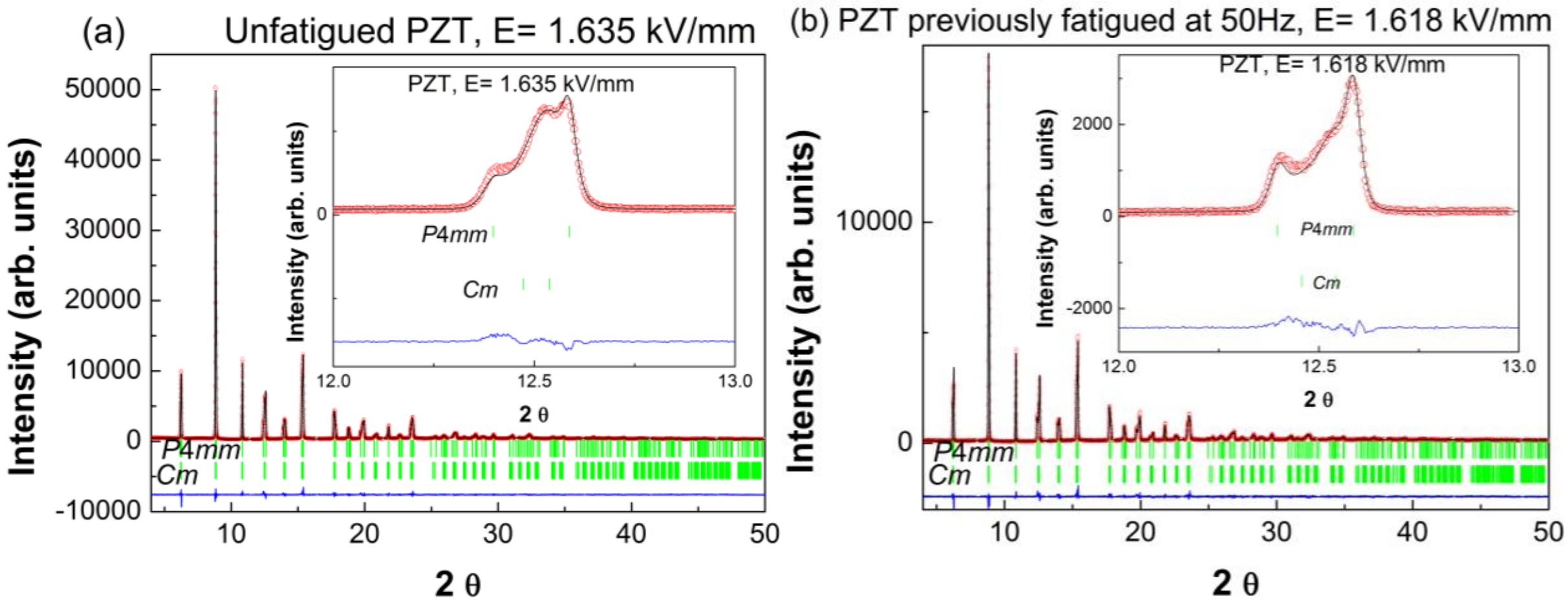

FIG. 2. Rietveld refinements for an unfatigued and a fatigued PZT. (a), unfatigued PZT at 1.635 kV/mm and (b) PZT at 1.618 kV/mm previously fatigued at 50 Hz kV/mm. The difference plot (blue) has the same scale. The vertical green tickmarks indicate the calculated positions of the *P*4*mm* phase (top) and *Cm* phase reflections (bottom). The inset corresponds to an enlargement of the $002_c$ reflection and the high-2θ range.

| | Unfatigued PZT | | PZT previously fatigued at 50 Hz | |
|---|---|---|---|---|
| Field increase | 0 kV/mm | | 0 kV/mm | |
| | *P*4*mm* | *Cm* | *P*4*mm* | *Cm* |
| *A* | 4.03645 (3) | 5.73082 (8) | 4.03641 (3) | 5.72980 (12) |
| *B* | | 5.72465 (10) | | 5.72645 (15) |
| *C* | 4.10067 (6) | 4.08035 (8) | 4.10052 (6) | 4.08194 (10) |
| $\beta$ | | 90.3167 (9) | | 90.310 (1) |
| | | | | |
| *x*-Zr, Ti | | 0.4876 (11) | | 0.4852 (17) |
| *z*-Zr, Ti | 0.5414 (12) | 0.5498 (11) | 0.5410 (13) | 0.5477 (15) |
| $B_{iso}$-Pb | 1.335 (33) | 2.37 (3) | 1.15 (4) | 1.82 (4) |
| *x*-O(1) | | 0.4430 (38) | | 0.4484 (50) |
| *z*-O(1) | 0.5614 (38) | 0.0323 (54) | 0.5707 (34) | 0.0502 (66) |
| *x*-O(2) | | 0.1974 (27) | | 0.1880 (34) |
| *y*-O(2) | | 0.2469 (23) | | 0.2472 (30) |
| *z*-O(2) | 0.0834 (42) | 0.5864 (26) | 0.0825 (45) | 0.5879 (35) |
| Phase Fraction | 35.7 (5) | 64.4 (6) | 42.5 (7) | 57.5 (8) |
| Bragg R-factor | 5.57 | 3.55 | 6.88 | 5.05 |

| | Unfatigued PZT | | PZT previously fatigued at 50 Hz | |
|---|---|---|---|---|
| Field increase | 0.5694 kV/mm | | 0.5176 kV/mm | |
| | *P*4*mm* | *Cm* | *P*4*mm* | *Cm* |
| *A* | 4.03666 (3) | 5.73042 (8) | 4.03634 (3) | 5.72921 (12) |
| *B* | | 5.72386 (10) | | 5.72493 (16) |
| *C* | 4.10154 (5) | 4.08171 (10) | 4.10130 (5) | 4.08293 (10) |
| $\beta$ | | 90.3223 (8) | | 90.3150 (13) |
| | | | | |
| *x*-Zr, Ti | | 0.4861 (10) | | 0.4848 (17) |
| *z*-Zr, Ti | 0.5421 (11) | 0.5486 (11) | 0.5393 (13) | 0.5477 (15) |
| $B_{iso}$-Pb | 1.39 (3) | 2.44 (3) | 1.21 (4) | 1.83 (4) |
| *x*-O(1) | | 0.4429 (41) | | 0.4455 (53) |
| *z*-O(1) | 0.5711 (30) | 0.0330 (57) | 0.5722 (32) | 0.0503 (68) |
| *x*-O(2) | | 0.1950 (28) | | 0.1860 (35) |
| *y*-O(2) | | 0.2436 (24) | | 0.2449 (32) |
| *z*-O(2) | 0.0824 (37) | 0.5853 (27) | 0.0783 (43) | 0.5827 (36) |
| Phase Fraction | 39.5 (5) | 60.5 (6) | 44.8 (8) | 55.2 (8) |
| Bragg R-factor | 5.43 | 3.87 | 6.76 | 4.89 |

| | Unfatigued PZT | | PZT previously fatigued at 50 Hz | |
|---|---|---|---|---|
| Field increase | 0,9988 kV/mm | | 1 kV/mm | |
| | *P*4*mm* | *Cm* | *P*4*mm* | *Cm* |
| *A* | 4.03728 (3) | 5.73071 (10) | 4.03669 (3) | 5.73020 (4) |
| *B* | | 5.72231 (11) | | 5.72154 (16) |
| *C* | 4.10164 (5) | 4.08114 (8) | 4.10128 (5) | 4.08037 (11) |
| $\beta$ | | 90.3122 (10) | | 90.3055 (15) |
| | | | | |
| *x*-Zr, Ti | | 0.4866 (12) | | 0.4848 (18) |
| *z*-Zr, Ti | 0.5424 (10) | 0.5468 (11) | 0.5422 (11) | 0.5432 (16) |
| $B_{iso}$-Pb | 1.30 (3) | 2.61 (3) | 1.21 (3) | 2.02 (2) |
| *x*-O(1) | | 0.4479 (45) | | 0.4574 (55) |
| *z*-O(1) | 0.5803 (24) | 0.0352 (60) | 0.5768 (27) | 0.0563 (67) |
| *x*-O(2) | | 0.1784 (27) | | 0.1784 (34) |
| *y*-O(2) | | 0.2349 (23) | | 0.2331 (30) |
| *z*-O(2) | 0.0866 (34) | 0.5706 (28) | 0.0832 (38) | 0.5835 (35) |
| Phase Fraction | 41.7 (5) | 58.3 (6) | 48.9 (8) | 51.1 (8) |
| Bragg R-factor | 6.28 | 4.17 | 7.15 | 5.22 |

| | Unfatigued PZT | | PZT previously fatigued at 50 Hz | |
|---|---|---|---|---|
| Field increase | 1.555 kV/mm | | 1.554 kV/mm | |
| | *P4mm* | *Cm* | *P4mm* | *Cm* |
| *A* | 4.03725 (4) | 5.73486 (9) | 4.03742 (4) | 5.73170 (13) |
| *B* | | 5.72827 (12) | | 5.72812 (16) |
| *C* | 4.09899 (7) | 4.07484 (10) | 4.09871 (7) | 4.07916 (10) |
| *β* | | 90.3070 (9) | | 90.2809 (15) |
| | | | | |
| *x*-Zr, Ti | | 0.4829 (10) | | 0.4829 (17) |
| *z*-Zr, Ti | 0.5419 (14) | 0.5532 (11) | 0.5443 (14) | 0.5498 (14) |
| $B_{iso}$-Pb | 1.21 (4) | 2.33 (2) | 1.06 (4) | 1.68 (3) |
| *x*-O(1) | | 0.4580 (41) | | 0.4441 (48) |
| *z*-O(1) | 0.5730 (37) | 0.0280 (58) | 0.5733 (38) | 0.0450 (64) |
| *x*-O(2) | | 0.1869 (26) | | 0.1865 (32) |
| *y*-O(2) | | 0.2440 (21) | | 0.2540 (28) |
| *z*-O(2) | 0.0927 (46) | 0.5982 (26) | 0.0860 (50) | 0.6047 (31) |
| Phase Fraction | 26.7 (4) | 73.3 (6) | 36.4 (7) | 63.6 (9) |
| Bragg R-factor | 5.69 | 3.03 | 6.4 | 4.62 |

| | Unfatigued PZT | | PZT previously fatigued at 50 Hz | |
|---|---|---|---|---|
| Field increase | 2.01 kV/mm | | 2 kV/mm | |
| | *P4mm* | *Cm* | *P4mm* | *Cm* |
| *A* | 4.03614 (3) | 5.73657 (9) | 4.03742 (5) | 5.7327 (2) |
| *B* | | 5.72954 (12) | | 5.7285 (2) |
| *C* | 4.09632 (6) | 4.07156 (9) | 4.09658 (9) | 4.0760 (12) |
| *β* | | 90.2878 (11) | | 90.2394 (16) |
| | | | | |
| *x*-Zr, Ti | | 0.4827 (11) | | 0.4825 (17) |
| *z*-Zr, Ti | 0.5404 (14) | 0.5522 (12) | 0.5424 (17) | 0.5479 (15) |
| $B_{iso}$-Pb | 1.26 (4) | 2.27 (2) | 0.82 (5) | 1.70 (3) |
| *x*-O(1) | | 0.4582 (40) | | 0.4462 (46) |
| *z*-O(1) | 0.5747 (37) | 0.0311 (56) | 0.5654 (56) | 0.0344 (65) |
| *x*-O(2) | | 0.1735 (24) | | 0.1930 (32) |
| *y*-O(2) | | 0.2371 (20) | | 0.2606 (28) |
| *z*-O(2) | 0.0829 (31) | 0.5862 (26) | 0.1016 (66) | 0.6091 (30) |
| Phase Fraction | 27.9 (5) | 72.1 (6) | 27.5 (7) | 72.5 (9) |
| Bragg R-factor | 5.64 | 2.81 | 7.55 | 4.56 |

Table 1: Cell parameters, fractional atomic coordinates, isotropic (iso) atomic displacement parameters ($Å^2$) and agreement factors (%) for monoclinic and tetragonal PZT as a function of *E* for unfatigued sample and sample previously fatigued at 50 Hz respectively. Due to the use of hard X-ray photons, only the lead isotropic displacement parameters were refined. *Cm*: site occupancies Pb, Zr/ Ti, O(1) 2*a* (*x*,0,*z*), O(2) 4*b* (*x*, *y* , *z*). Origin at Pb position.

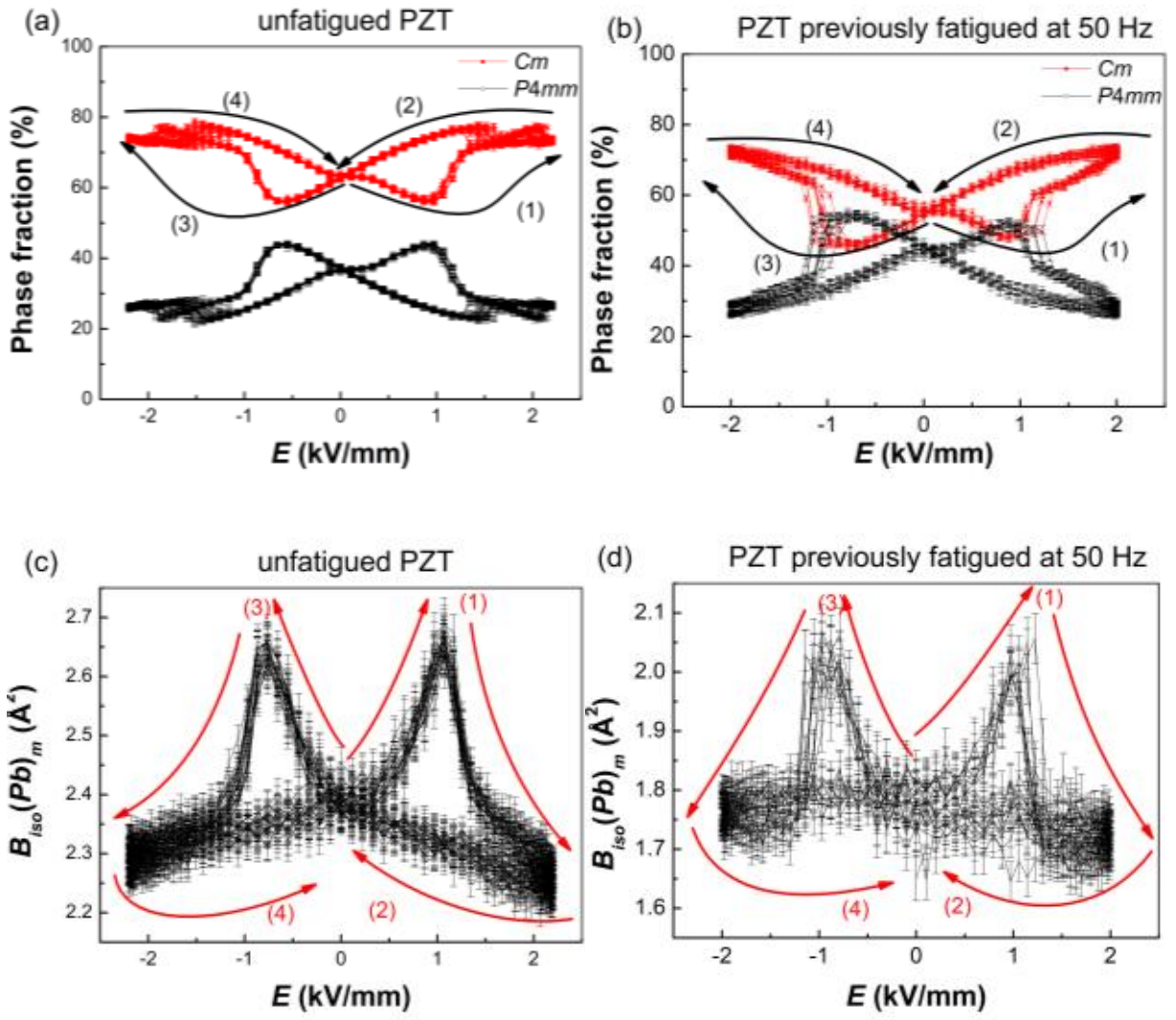

FIG. 3. Structural dependence of PZT sample as a function of the applied electric field obtained by Rietveld refinements. Monoclinic-Tetragonal phase ratio (%) for (a) the unfatigued and (b)the fatigued sample. Isotropic mean square displacement of the lead atom in the monoclinic phase $B_{iso}$ $(Pb)_{\mathrm{m}}$ for (c) the unfatigued and (d) the fatigued sample. Hysteresis cycles (HC) electric field values range (1-4) are indicated.

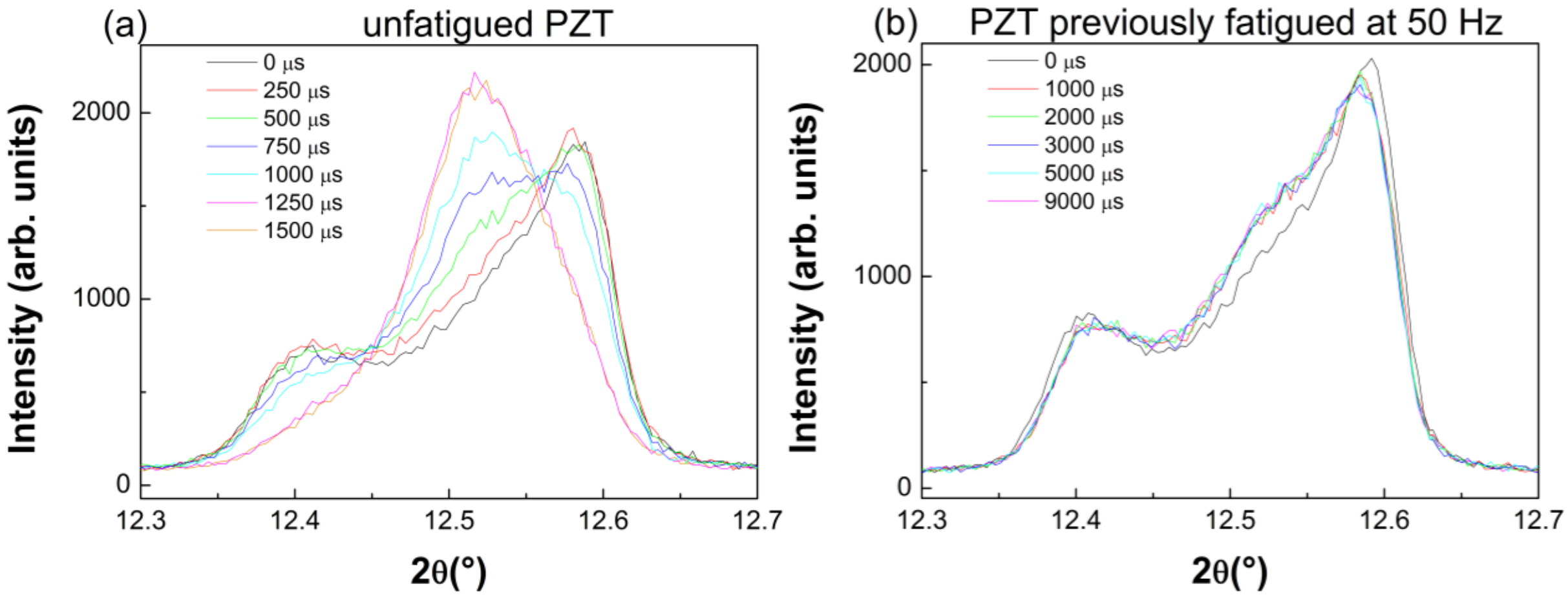

FIG. 4. Rectangular on-off stroboscopic pump-probe measurements of PZT. (a) Unfatigued pre-cycled sample; the sample was pre-cycled with $10^5$ cycles to assure a reproducible reaction throughout the whole experiment. (b) PZT fatigued with 50 Hz. In this setup the maximum field of 2.0 kV/mm is applied immediately and complete diffraction patterns are measured stroboscopically. The modulation of the field follows a rectangular on-off function with 50 Hz.